\documentclass[final,times,twocolumn]{elsarticle}
\usepackage{graphicx,amssymb,amsmath}
\usepackage{multirow,dcolumn,bm,latexsym,soul}
\newcolumntype{K}[1]{>{\centering\arraybackslash}p{#1}}

\journal{Physics Letters B}
\begin{document}
\newcommand{\be}{\begin{equation}}
\newcommand{\ee}{\end{equation}}
\newcommand{\bq}{\begin{eqnarray}}
\newcommand{\eq}{\end{eqnarray}}

\begin{frontmatter}

\title{Fine-structure constant constraints on late-time dark energy transitions}
\author[inst1,inst2]{C. J. A. P. Martins\corref{cor1}}\ead{Carlos.Martins@astro.up.pt}
\author[inst3]{M. Prat Colomer}\ead{maria.prat@estudiant.upc.edu}
\address[inst1]{Centro de Astrof\'{\i}sica da Universidade do Porto, Rua das Estrelas, 4150-762 Porto, Portugal}
\address[inst2]{Instituto de Astrof\'{\i}sica e Ci\^encias do Espa\c co, CAUP, Rua das Estrelas, 4150-762 Porto, Portugal}
\address[inst3]{Centre de Formaci\'o Interdisciplin\`aria Superior, Universitat Polit\`ecnica de Catalunya, C/Pau Gargallo 14, 08028 Barcelona, Spain}
\cortext[cor1]{Corresponding author}

\begin{abstract}
We use recent astrophysical and local tests of the stability of the fine-structure constant, $\alpha$, to constrain a particular phenomenological but physically motivated class of models in which the dark energy equation of state can undergo a rapid transition at low redshifts, perhaps associated with the onset of the acceleration phase. We set constraints on the phenomenological parameters describing such possible transitions, in particular improving previous constraints (which used only cosmological data) on the present-day value of the dark energy equation of state in these models. We specifically quantify how these constraints are improved by the addition of the $\alpha$ measurements. We find no evidence for a transition associated with the onset of acceleration. In this model the $\alpha$ measurements lead to a bound on Weak Equivalence Principle violations of $\eta<4\times10^{-15}$ (at $68.3\%$ confidence level), improving on the recent MICROSCOPE bound by about a factor of three.
\end{abstract}


\end{frontmatter}


\section{Introduction}

Mapping the recent dynamics of the universe is among the most pressing tasks of observational cosmology. While there is compelling evidence for its recent acceleration \cite{SN1,SN2}, the properties and origin of the physical mechanism which is responsible for this---whether it's a cosmological constant or a dynamical degree of freedom---are still unknown. One observational approach consists of mapping the behaviour of dark energy as a function of redshift and trying to identify any dynamics \cite{Huterer}, and this is commonly done in terms of the dark energy equation of state. It is observationally clear that the present-day value of the dark energy equation of state, $w_0$, must be very close to $-1$ \cite{Ade:2015rim} (though these constraints depend on assumptions about the underlying model), but the constraints at higher redshifts are weaker.

In recent work \cite{MPC} we used Type Ia Supernova data \cite{Suzuki} and Hubble parameter measurements \cite{Farooq} to constrain models where the dark energy equation of state can undergo a late-time phase transition. Physical mechanisms that may lead to such transitions include topological defects \cite{Hill:1988vm}, vacuum metamorphosis \cite{Parker1,Parker2}, and scalar field models with a non-canonical kinetic term \cite{Mortonson}. Here we improve upon our previous analysis in several ways. First, we use the latest Pantheon supernova sample \cite{Pantheon} which spans a larger redshift range. Second, the present-day value of the Hubble parameter is always analytically marginalized; while the analysis in \cite{MPC} suggests that keeping it fixed has a relatively small impact, marginalization is a safer option given the ongoing controversy regarding its value \cite{HRiess,HPlanck}. And thirdly (and most significantly), we include in the analysis recent astrophysical and local tests of the stability of the fine-structure constant $\alpha$---see \cite{ROPP} for a recent review. 

In realistic dynamical dark energy models the new degree of freedom (presumably a scalar field) is expected to couple to the rest of the model, unless a yet-unknown symmetry is postulated to suppress these couplings \cite{Carroll}. In particular, a coupling of the field to the electromagnetic sector will lead to spacetime variations of the fine-structure constant $\alpha$, whose measurements can therefore be used to constrain dark energy \cite{Parkinson,NunLid,Doran}. Moreover, this coupling will also lead to a violation of the Weak Equivalence Principle (henceforth WEP). Thus measurements of $\alpha$ have cosmological implications that go beyond the mere fundamental nature of the tests themselves. In this work we will therefore make the 'minimal' (parsimonious) assumption that the dynamical scalar field responsible for the dark energy also leads to a redshift dependence of $\alpha$. Moreover, in our statistical analysis we will also use as prior the recent MICROSCOPE bound on Weak Equivalence Principle violations \cite{Touboul}, and quantify how these $\alpha$ measurements improve on it.

We will show that despite enlarging the parameter space with respect to earlier analyses (and including both cosmological and particle physics free parameters) we will still obtain stringent constraints, since the additional astrophysical and local tests included in the analysis are themselves quite strong.


\section{Equation of state parametrisations and cosmological constraints}

A transition in the dark energy equation of state can generically be described with four free parameters: the values of the equation of state itself at early and late times (respectively denoted $w_i$ and $w_f$), plus the characteristic redshift at which the transition occurs, $z_t$, and a transition width $\Delta$ (describing whether the transition is fast or slow). The present-day value of the dark energy equation of state, $w_0$, can then be expressed in terms of these four parameters, or alternatively replace one of them.

Several phenomenological parametrisations of the dark energy equation of state have been proposed and studied, which allow for a fast transition, for example \cite{Bassett,Corasaniti,Linder:2005ne,Lazkoz,Jaber,Durrive}; these were compared in \cite{MPC} and found to be qualitatively similar. In what follows we will use that of Linder \& Huterer \cite{Linder:2005ne}, since it leads to an analytic form for the Friedmann equation (which somewhat simplifies the statistical analysis). We further assume that the late-time value corresponds to a cosmological constant ($w_f=-1$) and that models are canonical, having $w(z)\ge-1$. In this case the dark energy equation of state has the form
\be
w(z)=\frac{w_i(1+w_0)(1+z)^{1/\Delta}-(w_i-w_0)}{(1+w_0)(1+z)^{1/\Delta}+(w_i-w_0)}\,; \label{eos4}
\ee
note that if either $w_0=-1$ or $w_i=-1$ this parametrisation reduces to $w(z)=-1$ throughout. The characteristic redshift of the transition is
\be
z_t=\left(\frac{w_i-w_0}{1+w_0}\right)^\Delta-1\,. \label{Modell23}
\ee
Note that one can make choices of parameters for which one would obtain a negative transition redshift, implying that in such cases the effective epoch of transition is in the future. However, one finds that such choices are observationally excluded. We further assume flat Friedmann-Lema\^{\i}tre-Robertson-Walker models, so that the present-day values of the matter and dark energy densities (expressed as a function of the critical density) satisfy $\Omega_m+\Omega_{dark}=1$. We neglect the contribution of the radiation density, since we will only be concerned with low-redshift data. Then the Friedmann equation is
\be\label{hubb4}
\frac{H^2}{H_0^2}=\Omega_m(1+z)^3+(1-\Omega_m)\left[\frac{1+w_0}{1+w_i}(1+z)^{1/\Delta}+\frac{w_i-w_0}{1+w_i}\right]^{3\Delta(1+w_i)}\,.
\ee

We start by revisiting \cite{MPC} and constrain this model using low-redshift background cosmology data, specifically the recent catalogue of Type Ia Pantheon supernovas \citep{Pantheon} (while \cite{MPC} used the earlier Union2.1 catalogue \cite{Suzuki}) together with the compilation of measurements of the Hubble parameter by \citet{Farooq}. The Hubble constant is always analytically marginalized, following the procedure detailed in \cite{Basilakos}.

We carry out a standard statistical likelihood analysis, assuming flat logarithmic prior on $(1+w_0)$ and $(1+w_i)$, specifically $\log_{10}{(1+w_0)}=[-2,-0.5]$ and $\log_{10}{(1+w_i)}=[-2,0]$, and a uniform prior on the transition width, $\Delta=[0,1]$. Our choice of priors for $w_0$ and $w_i$ is motivated by the fact that (as will be seen shortly) the data prefers their values to be very close to those of a cosmological constant, and that for values smaller than $\log_{10}{(1+w)}=-2$ the phase space volume has almost the same likelihood. (Note that a choice of a logarithmic prior for a parameter whose value is near zero may affect the resulting posterior, due to the effect of prior volume weighting.)

To facilitate the comparison with our earlier work \cite{MPC}, we fix the present-day value of the matter density to be $\Omega_m=0.3$, in agreement with both CMB and low-redshift data \cite{HPlanck,Pantheon,DES}. Note that one of our goals is to determine how well these models can be constrained by low-redshift data alone, and that the inclusion of the astrophysical data is not expected to significantly affect the preferred value of $\Omega_m$ \cite{ROPP}. On the one hand, allowing $\Omega_m$ to vary would weaken our constraints, but on the other hand they could be made more stringent by adding, e.g., Cosmic Microwave Background data. Thus overall the fixed matter density is a simplifying assumption which should not sigificantly bias our results.

As a further check of the impact of our assumptions on the final results, we also repeat the analysis for two more specific cases of this parametrisation. The first particular case corresponds to fixing $w_i=0$; this will ensure a transition (from $w=0$ to about $w=-1$), which the data will push to high redshifts \cite{MPC}, though note that this choice implies that at early times we will have an additional matter-like component. In other words, such models have a non-zero amount of early dark energy \cite{Wetterich}, which is constrained by Cosmic Microwave Background experiments such as Planck to sub-percent level \cite{Ade:2015rim}. The second particular case corresponds to fixing $\Delta=1/3$, as recently done in \cite{Durrive}, on the grounds that this is the value for which this parametrisation best approximates the dynamics of a specific sub-class of quintessence models known as scaling freezing models.

\begin{figure}
\begin{center}
\includegraphics[width=3.2in,keepaspectratio]{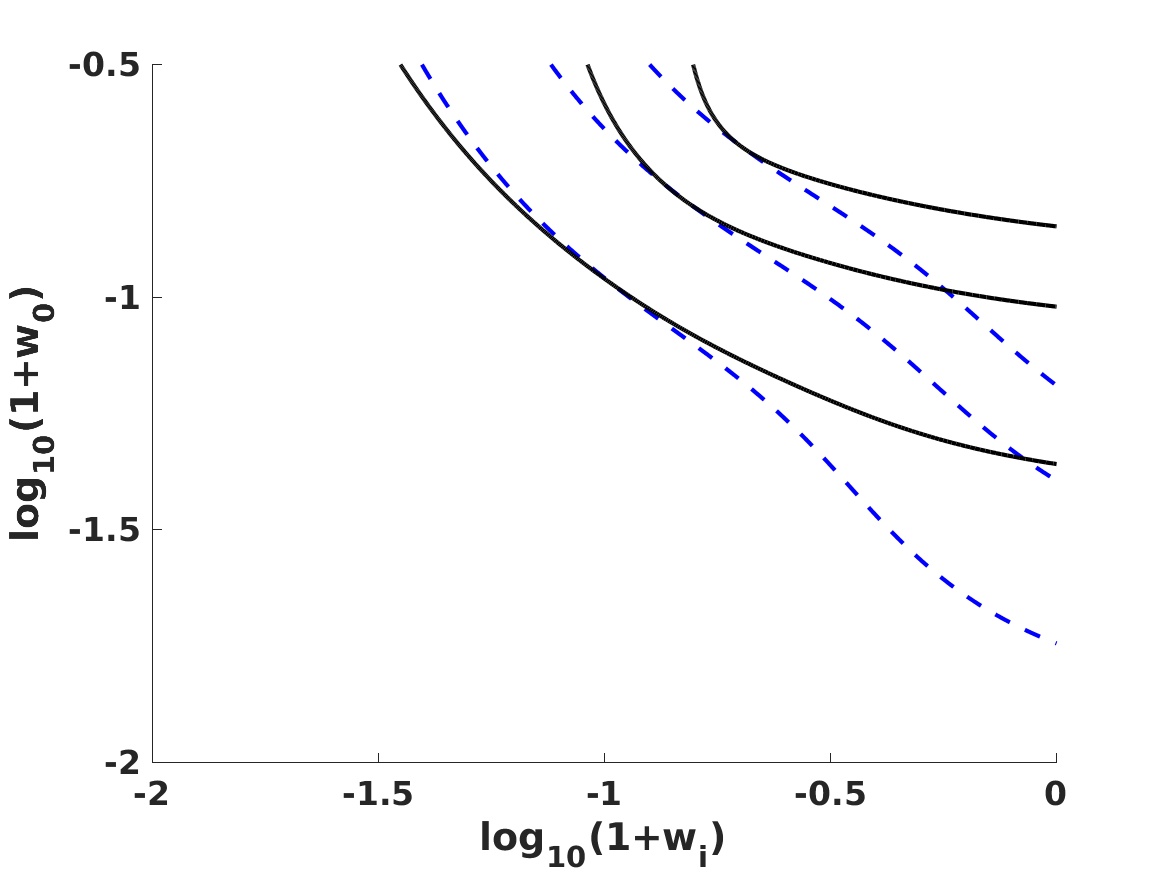}
\includegraphics[width=3.2in,keepaspectratio]{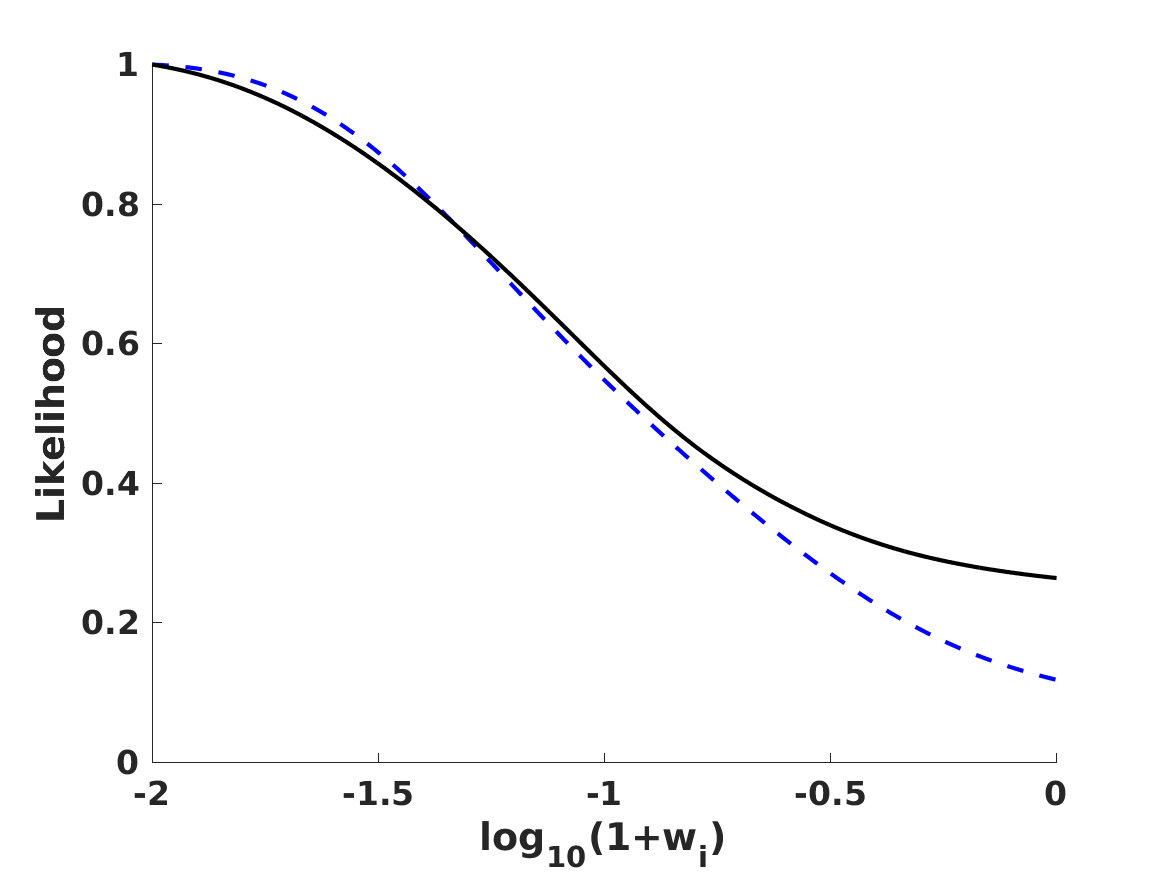}
\includegraphics[width=3.2in,keepaspectratio]{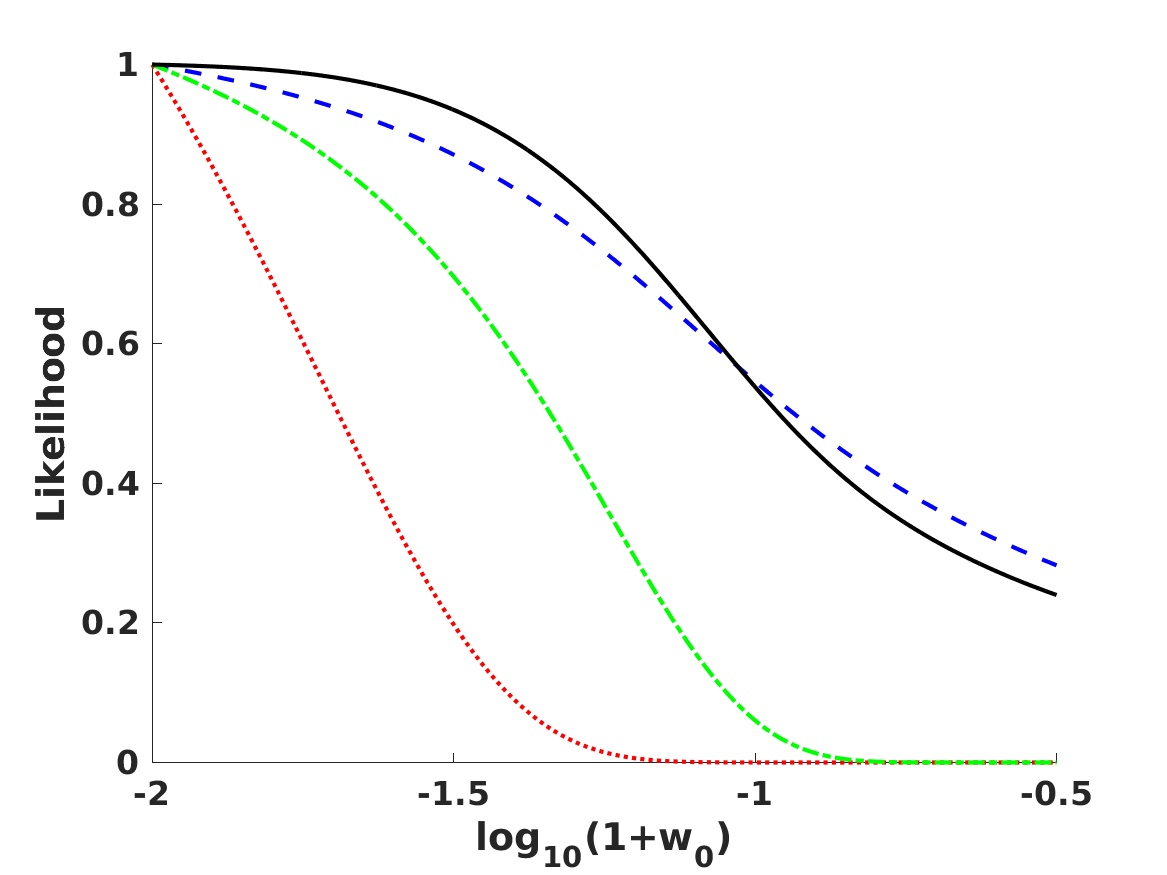}
\end{center}
\caption{\label{fig1}Constraints on $w_i$ and $w_0$ from our cosmological data, with the remaining model parameters marginalized. The first panel shows one, two and three sigma constraints on the $w_i-w_0$ plane, while the other two panels show the posterior likelihoods for each of the parameters. The red curves correspond to $w_i=0$ and $\Delta=1/3$; green curves to $w_i=0$ and $\Delta$ allowed to vary; blue curves to $\Delta=1/3$ and $w_i$ allowed to vary; black curves to the case with both parameters allowed to vary (as described in the text).}
\end{figure}

\begin{table}
\begin{center}
\caption{One sigma ($68.3\%$ confidence level) upper bounds on the early and present values of the dark energy equation of state, $w_i$ and $w_0$, under the various assumptions described in the text. Note that the logarithms are base 10.}
\label{table1}
\begin{tabular}{| c  c | c | c |}
\hline
$w_i$ & $\Delta$ & $w_i$ & $w_0$ \\
\hline
Fixed & Fixed & N/A & $<-0.982$ \\
\hline
Free & Fixed & $<-0.919$ & $<-0.917$ \\
\hline
Fixed & Free & N/A & $<-0.962$ \\
\hline
Free & Free & $<-0.913$ & $<-0.915$ \\
\hline
\end{tabular}
\end{center}
\end{table}

Figure \ref{fig1}  and Table \ref{table1} summarize this analysis, which improves upon and updates that of \cite{MPC}. In all cases, the data strongly prefers a present equation of state close to a cosmological constant. With a fixed value $w_i=0$, the putative transition is pushed to high redshifts (into the matter era), while if $w_i$ is allowed to vary it is also constrained to be close to a cosmological constant, in which case there is effectively no transition (so the transition width $\Delta$ is unconstrained). The choice of a free or fixed $\Delta$ has a significant impact if $w_i=0$ (as expected, since in that case a transition is effectively imposed), but a much smaller one if $w_i$ is allowed to vary (since then there is no transition).


\section{Varying $\alpha$ and astrophysical constraints}

Dynamical scalar fields in an effective 4D field theory are naturally expected to couple to the rest of the theory, unless one is prepared to postulate that a (still unknown) symmetry suppresses these couplings \cite{Carroll,Dvali,Chiba}. We assume that our phenomenological parametrisation, discussed in the previous section, models the dynamics of a scalar field (denoted $\phi$) which couples to the electromagnetic sector via a gauge kinetic function $B_F(\phi)$

\begin{equation}
{\cal L}_{\phi F} = - \frac{1}{4} B_F(\phi) F_{\mu\nu}F^{\mu\nu}\,.
\end{equation}
This function can be assumed to be linear,
\begin{equation}
B_F(\phi) = 1- \zeta \kappa (\phi-\phi_0)\,,
\end{equation}
(where $\kappa^2=8\pi G$) since the absence of such a term would require a $\phi\to-\phi$ symmetry, but such any such symmetry must be broken throughout most of the cosmological evolution \cite{Dvali}. The dimensionless parameter $\zeta$ quantifies the strength of the coupling. With these assumptions one can explicitly relate the evolution of $\alpha$ to that of dark energy \cite{Erminia2,Pinho2} as follows
\begin{equation}
\frac{\Delta \alpha}{\alpha} \equiv \frac{\alpha-\alpha_0}{\alpha_0} =B_F^{-1}(\phi)-1=
\zeta \kappa (\phi-\phi_0) \,,
\end{equation}
and defining the fraction of the dark energy density (the ratio of the energy density of the scalar field to the total energy density, which also includes a matter component) as a function of redshift $z$ as
\begin{equation}
\Omega_\phi (z) \equiv \frac{\rho_\phi(z)}{\rho_{\rm tot}(z)} \simeq \frac{\rho_\phi(z)}{\rho_\phi(z)+\rho_m(z)} \,,
\end{equation}
where in the last step we have again neglected the contribution from the radiation density, the evolution of the scalar field can be expressed in terms of $\Omega_\phi$ and of the dark energy equation of state $w_\phi$ as
\begin{equation}\label{dynphi}
1+w_\phi = \frac{(\kappa\phi')^2}{3 \Omega_\phi} \,;
\end{equation}
here, consistently with the previous section, we have made the simplifying assumption of a canonical scalar field, and the prime denotes the derivative with respect to the logarithm of the scale factor. Putting the two together and expressing the evolution in terms of the redshift we finally obtain
\begin{equation} \label{eq:dalfa}
\frac{\Delta\alpha}{\alpha}(z) =\zeta \int_0^{z}\sqrt{3\Omega_\phi(z')\left[1+w_\phi(z')\right]}\frac{dz'}{1+z'}\,.
\end{equation}

Thus in these models the evolution of $\alpha$ can be expressed as a function of cosmological parameters plus the coupling $\zeta$, without explicit reference to dynamics of the underlying scalar field. In our analysis, the cosmological model is specified by Eqs. \ref{eos4} and \ref{hubb4}. Another important observable is the current drift rate of $\alpha$, which can be found, from Equation \ref{eq:dalfa}, to be
\begin{equation}
\label{eq: drift}
\left(\dfrac{1}{H} \dfrac{\dot{\alpha}}{\alpha}\right)_0 = -\zeta\sqrt{3 \Omega_{\phi 0} (1+w_0)}\,.
\end{equation}
Naturally this just depends on the present value of the dark energy equation of state, $w_0$, and vanishes for $w_0=-1$. This provides a first way to constrain these models using local experiments. A second one stems from the fact that a light scalar field such as we are considering inevitably couples to nucleons due to the $\alpha$ dependence of their masses, and therefore it mediates an isotope-dependent long-range force. This can be quantified through the dimensionless E\"{o}tv\"{o}s parameter $\eta$, which describes the level of violation of the WEP. One can show that in this class of canonical quintessence-type models the E\"{o}tv\"{o}s parameter and the dimensionless coupling $\zeta$ are simply related by \cite{Dvali,Chiba}
\begin{equation} \label{eq:eotvos}
\eta \approx 10^{-3}\zeta^2\,;
\end{equation}
therefore, local tests of the Equivalence Principle also constrain the dimensionless coupling parameter $\zeta$.

We can now repeat our statistical analysis including data from astrophysical (spectroscopic) tests of the stability of $\alpha$. (For a recent summary of the methodology behind these tests and their current observational status, see \cite{ROPP}.) We will use all the currently available measurements: the Webb {\it et al.} \cite{Webb} dataset (a large dataset of 293 archival data measurements), the smaller dataset of 21 dedicated measurements compiled in \cite{ROPP} (which are expected to have a better control of possible systematics), and the more recent measurements of \cite{Cooksey}. Additionally we use three other local constraints:
\begin{itemize}
\item Firstly, the Oklo natural nuclear reactor \cite{Oklo} provides a geophysical constraint on the value of $\alpha$
\begin{equation} \label{okloalpha}
\frac{\Delta\alpha}{\alpha} =0.005\pm0.061\, ppm\,,
\end{equation}
at an effective redshift $z=0.14$, under the further simplifying assumption that $\alpha$ is the only parameter than may have been different and all the remaining physics is unchanged.
\item Secondly, the atomic clocks laboratory bound of \cite{Rosenband} constrains the current drift of $\alpha$ to be
\begin{equation} \label{rosen}
\left(\frac{1}{H}\frac{\dot\alpha}{\alpha}\right)_0=-0.22\pm0.32\, ppm\,;
\end{equation} 
note that for convenience we use units of parts per million (ppm) throughout, both for values of $\alpha$ and for those of the coupling $\zeta$.
\item Thirdly, we will use the recently improved constraint on the  E\"{o}tv\"{o}s parameter obtained by the MICROSCOPE satellite \cite{Touboul}
\begin{equation} \label{micro}
\eta=(-0.1\pm1.3)\times10^{-14}\,.
\end{equation}
\end{itemize}

\begin{figure}
\begin{center}
\includegraphics[width=3.2in,keepaspectratio]{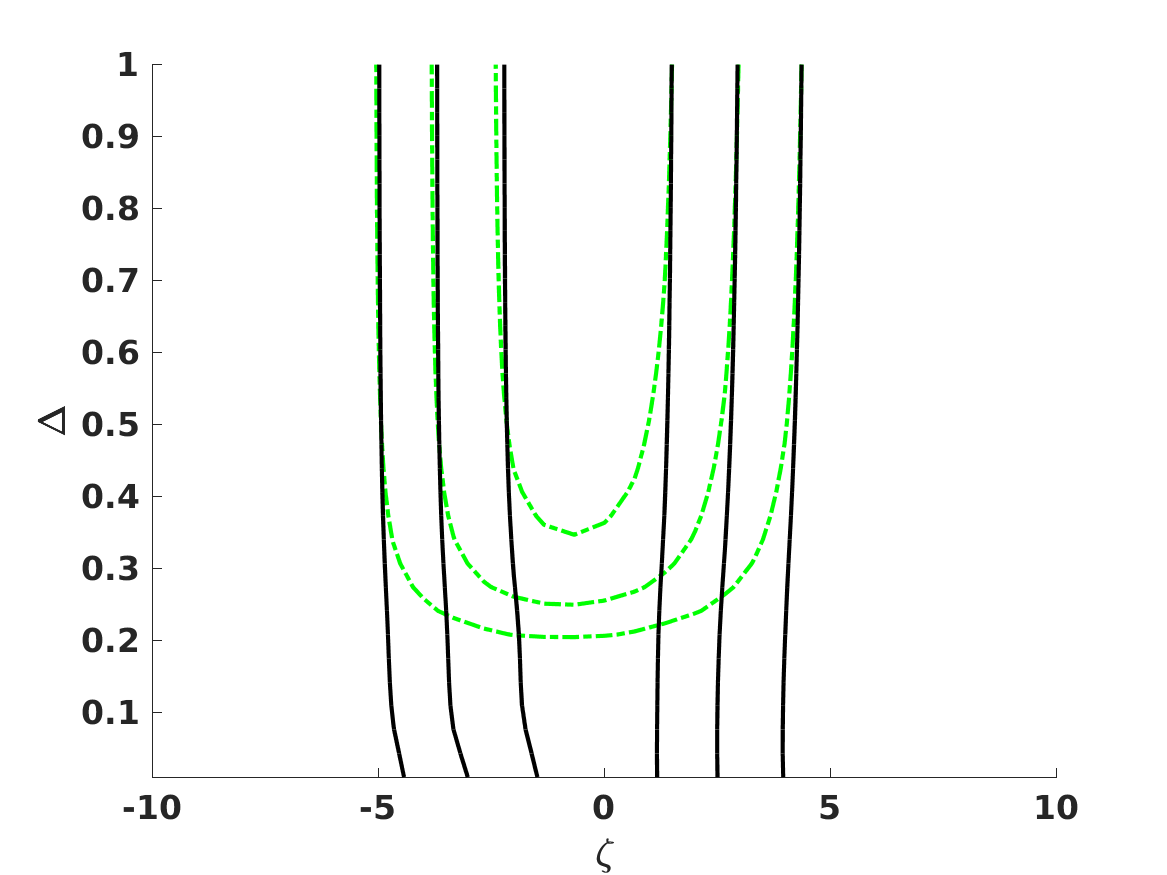}
\includegraphics[width=3.2in,keepaspectratio]{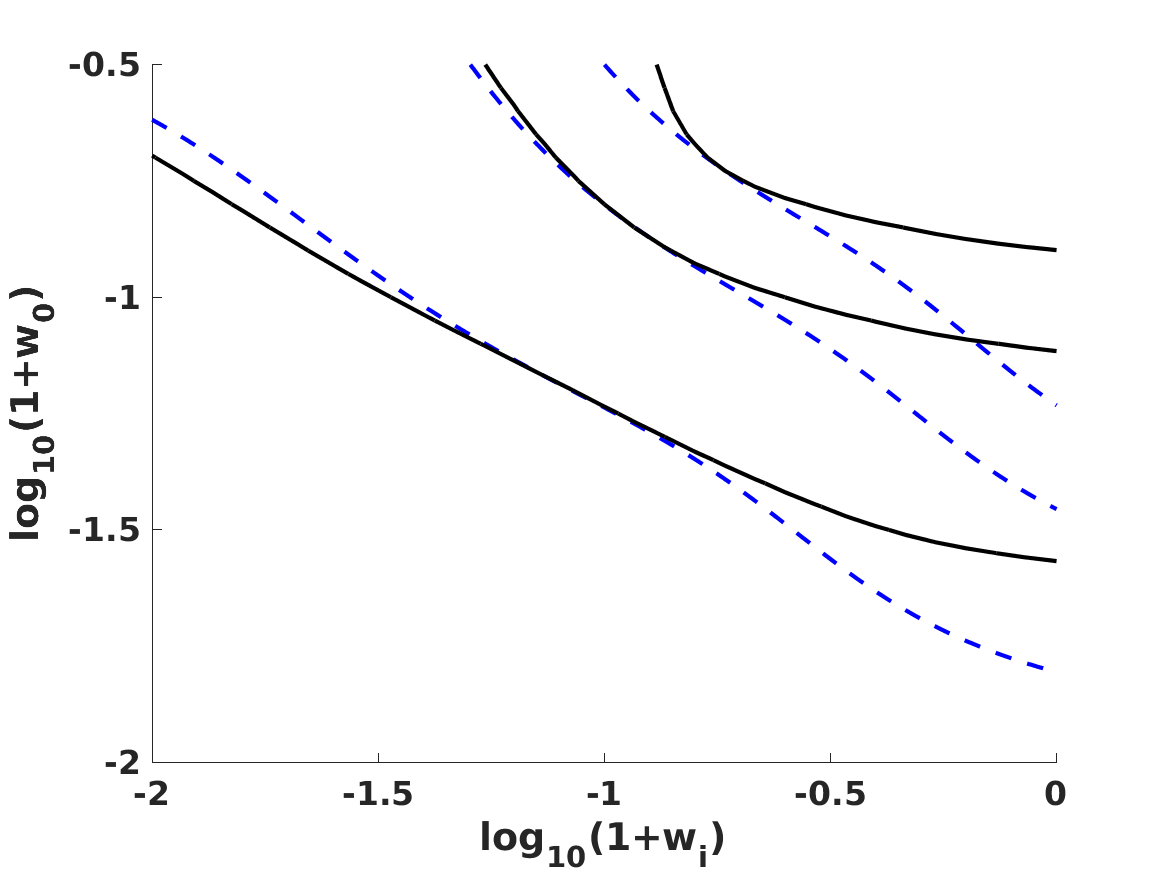}
\includegraphics[width=3.2in,keepaspectratio]{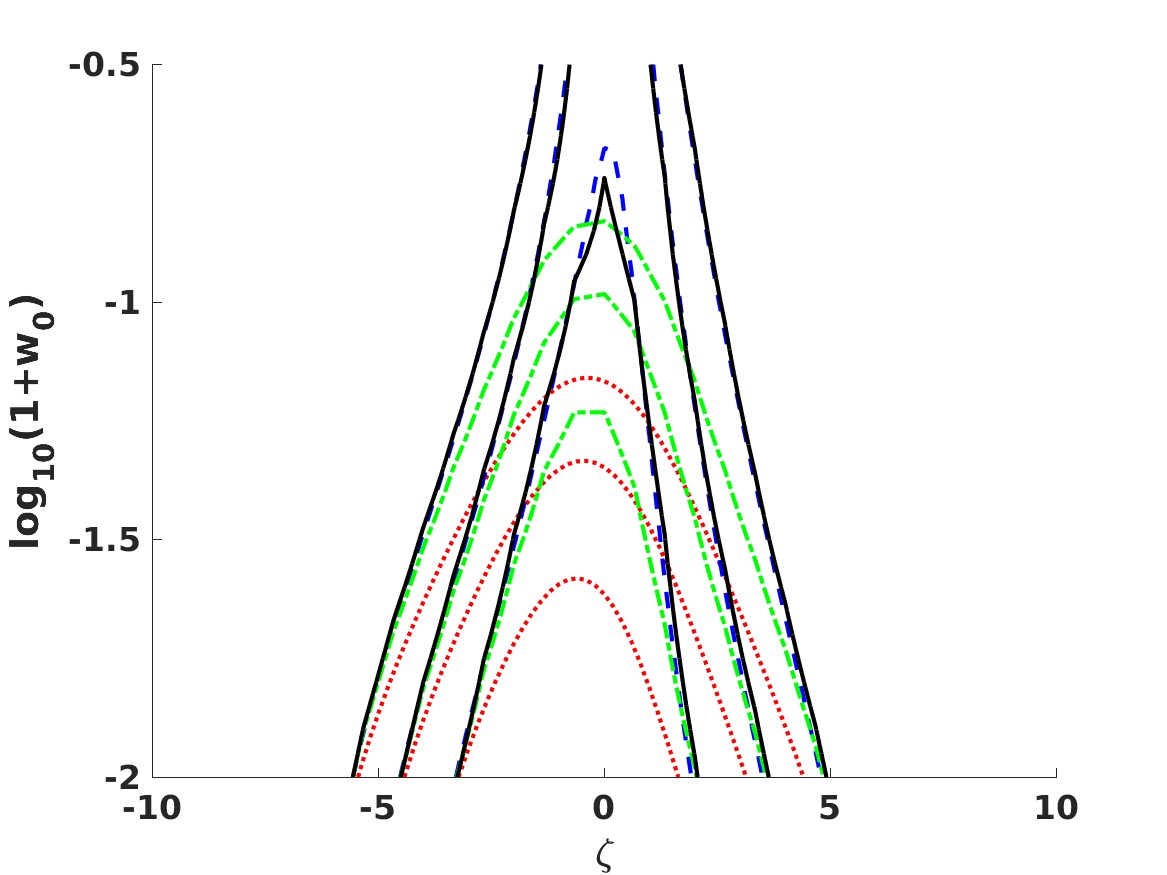}
\end{center}
\caption{\label{fig2}One, two and three sigma constraints on relevant two-dimensional planes from our full dataset, with remaining parameters marginalized. The red curves correspond to $w_i=0$ and $\Delta=1/3$; green curves to $w_i=0$ and $\Delta$ allowed to vary; blue curves to $\Delta=1/3$ and $w_i$ allowed to vary; black curves to the case with both parameters allowed to vary (as described in the text).}
\end{figure}

\begin{figure}
\begin{center}
\includegraphics[width=3.2in,keepaspectratio]{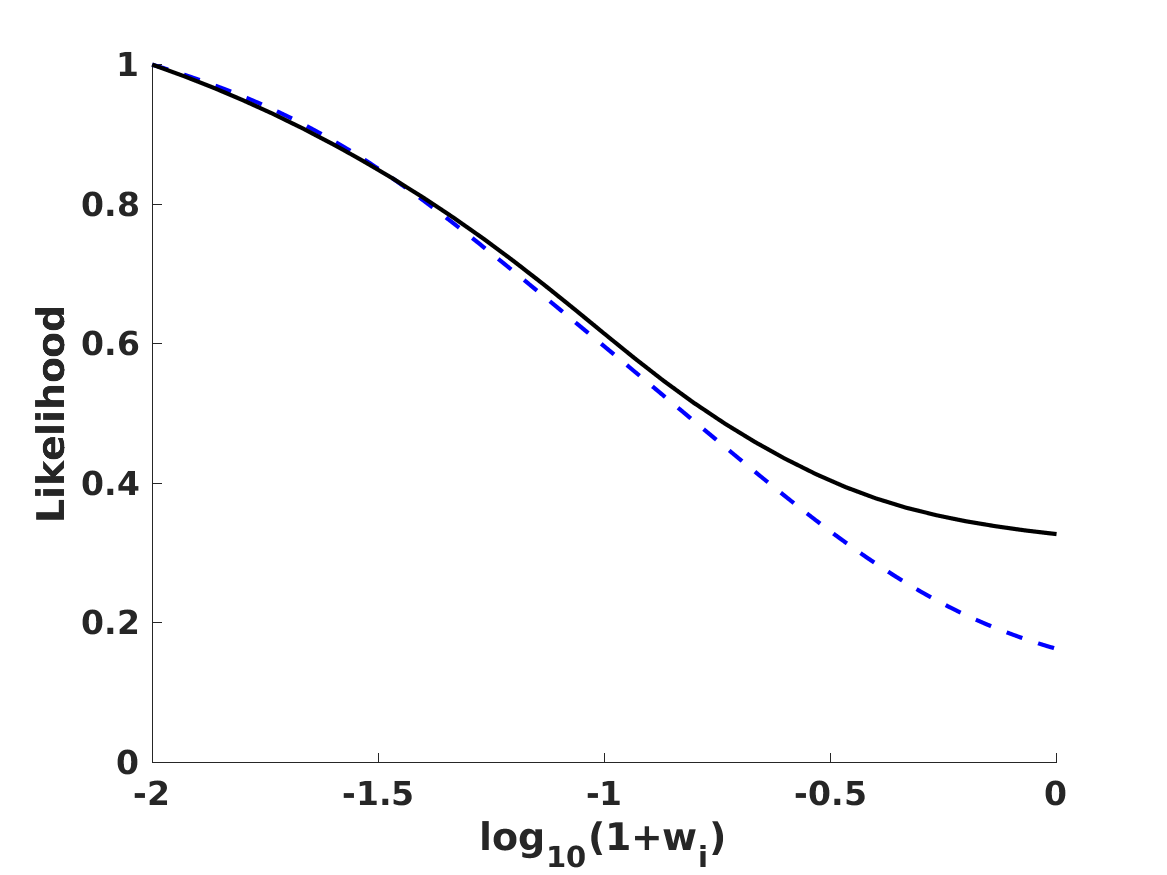}
\includegraphics[width=3.2in,keepaspectratio]{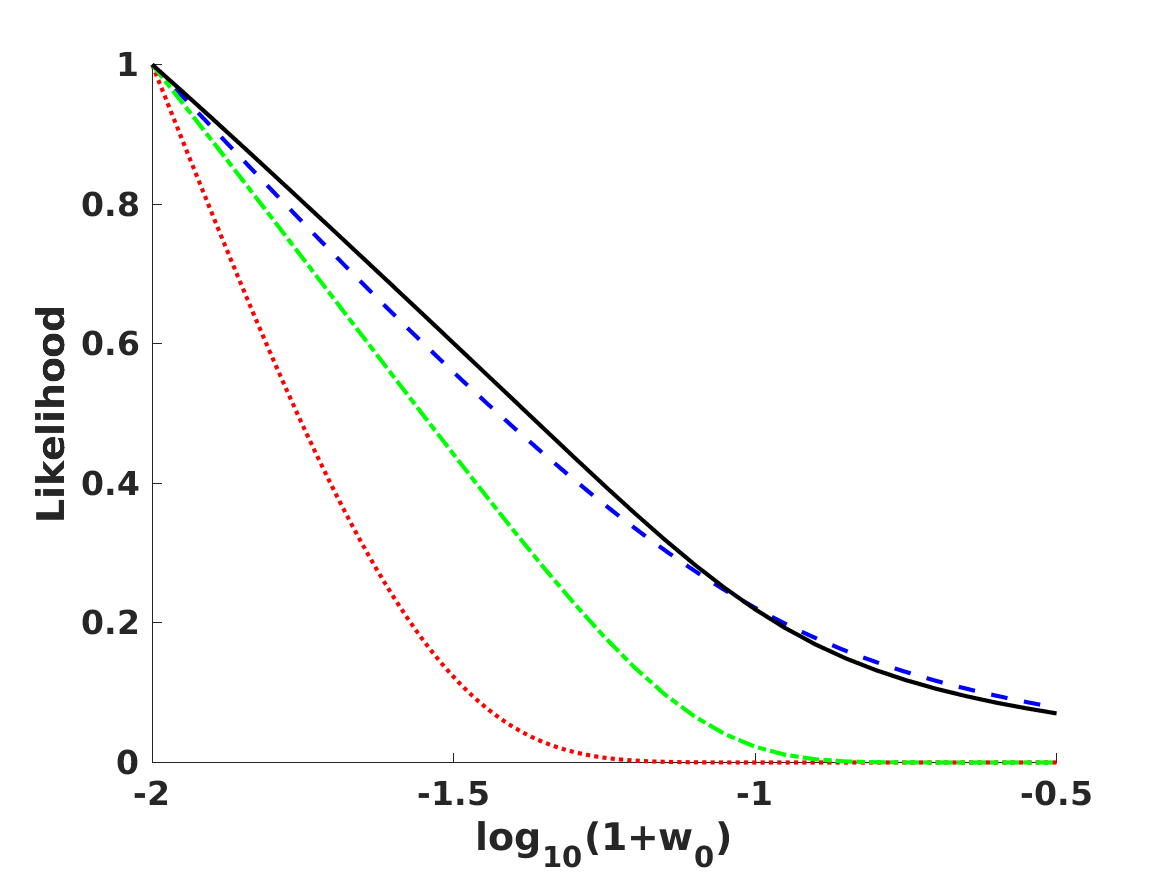}
\includegraphics[width=3.2in,keepaspectratio]{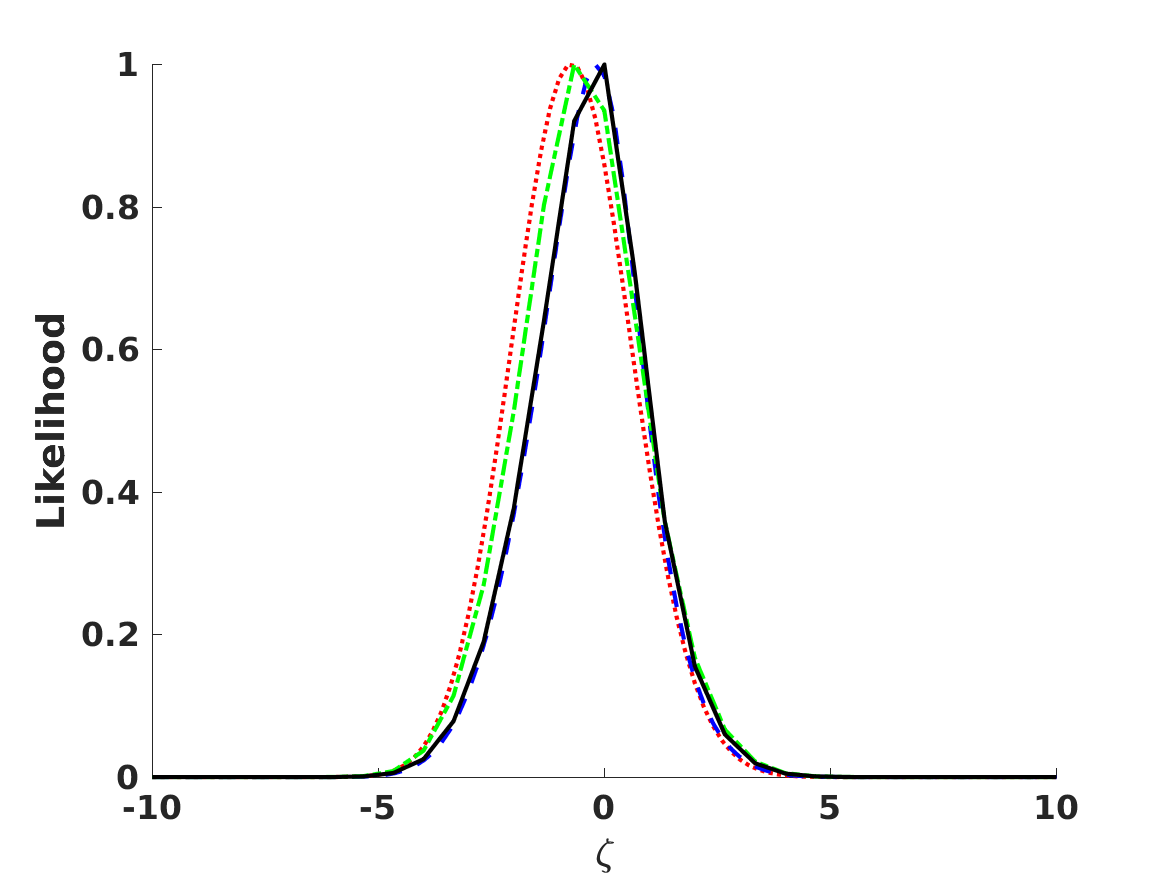}
\end{center}
\caption{\label{fig3}Posterior likelihoods for $w_i$,  $w_0$ and $\zeta$ from our full dataset, with remaining parameters marginalized. The red curves correspond to $w_i=0$ and $\Delta=1/3$; green curves to $w_i=0$ and $\Delta$ allowed to vary; blue curves to $\Delta=1/3$ and $w_i$ allowed to vary; black curves to the case with both parameters allowed to vary (as described in the text).}
\end{figure}

We keep the same assumptions on cosmological parameters as in the previous section, but enlarge the parameter space by including the coupling $\zeta$, for which we assume a flat prior  $\zeta=[-10,10]$ ppm. The results are summarized in Figures \ref{fig2} and \ref{fig3} and in Table \ref{table2}. Broadly speaking the constraints on $w_i$ become slightly weaker (as a result of the larger parameter space), but those on $w_0$ become stronger; the latter is due to the strong local constraints coming from atomic clocks and (indirectly) also the MICROSCOPE WEP bound. On the other hand the constraint on $\zeta$ is fairly robust to the different assumptions on $\Delta$ and $w_i$, both due to the tight local constraints and because cosmological data constrains $w(z)$ to be close to $w=-1$ throughout most of the redshift range of current astrophysical measurements.

\begin{table}
\begin{center}
\caption{One sigma ($68.3\%$ confidence level) upper bounds on the early and present values of the dark energy equation of state, $w_i$ and $w_0$, and one-sigma constraints on the coupling $\zeta$ (in parts per million units), under the various assumptions described in the text. Note that the logarithms are base 10.}
\label{table2}
\begin{tabular}{| c  c | c | c | c |}
\hline
$w_i$ & $\Delta$ & $w_i$ & $w_0$ & $\zeta$ (ppm) \\
\hline
Fixed & Fixed & N/A & $<-0.985$ & $-0.8\pm1.3$ \\
\hline
Free & Fixed & $<-0.905$ & $<-0.972$ & $-0.2\pm1.1$ \\
\hline
Fixed & Free & N/A & $<-0.978$ & $-0.7\pm1.4$ \\
\hline
Free & Free & $<-0.900$ & $<-0.968$ & $-0.1\pm1.3$ \\
\hline
\end{tabular}
\end{center}
\end{table}

Finally, having used Eq. \ref{eq:eotvos} to convert the MICROSCOPE constraint on the E\"{o}tv\"{o}s parameter into a constraint on the coupling $\zeta$ (effectively a further Gaussian prior on it), we can now use it again and convert the posterior likelihood on $\zeta$ from the combination of local, astrophysical and cosmological data into an indirect bound on the E\"{o}tv\"{o}s. Conservatively assuming the bound $|\zeta|<2$ ppm, we straightforwardly obtain
\begin{equation} \label{eq:eotvos2}
\eta < 4\times10^{-15}\,,
\end{equation}
which improves the MICROSCOPE bound by more than a factor of three. This is an indirect and somewhat model-dependent constraint, but nevertheless the model dependence is expected to be relatively mild---see \cite{Pinho2} for additional discussion of this point.

\begin{figure*}
\begin{center}
\includegraphics[width=3.2in,keepaspectratio]{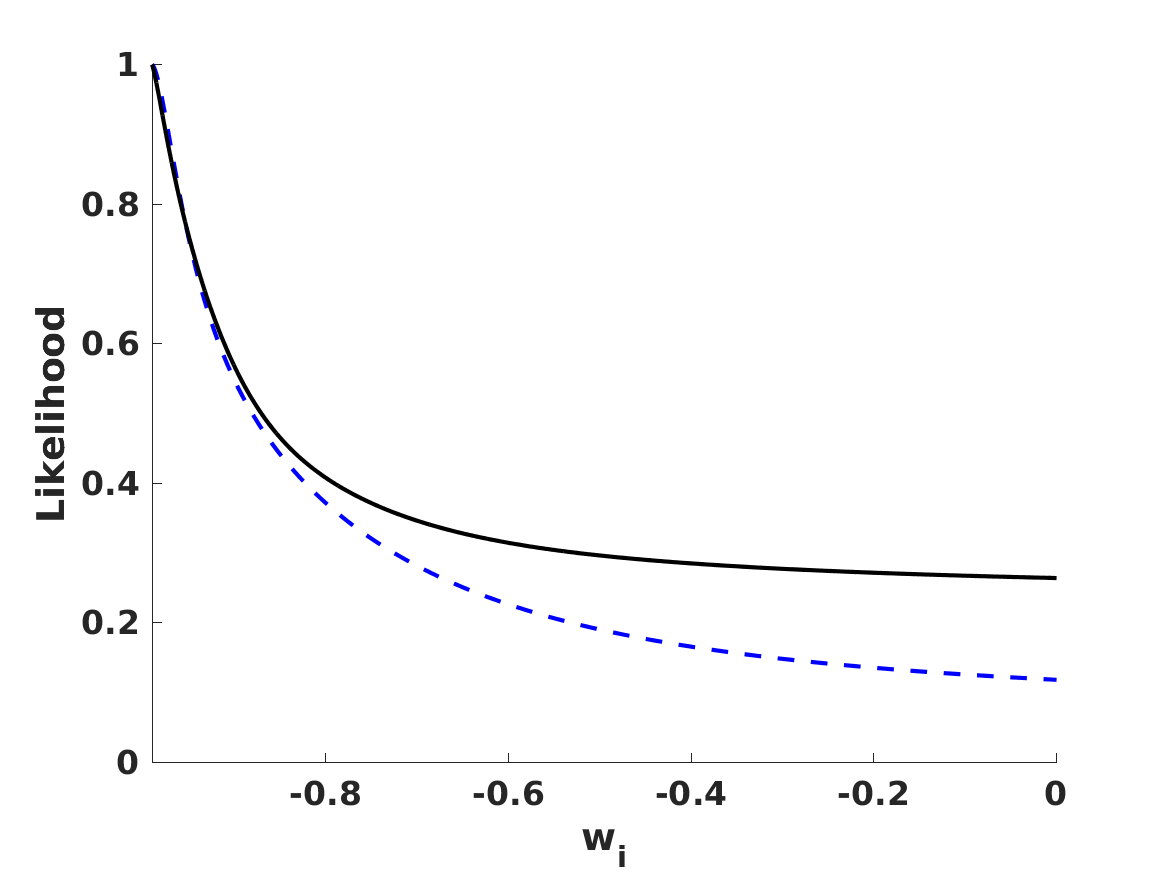}
\includegraphics[width=3.2in,keepaspectratio]{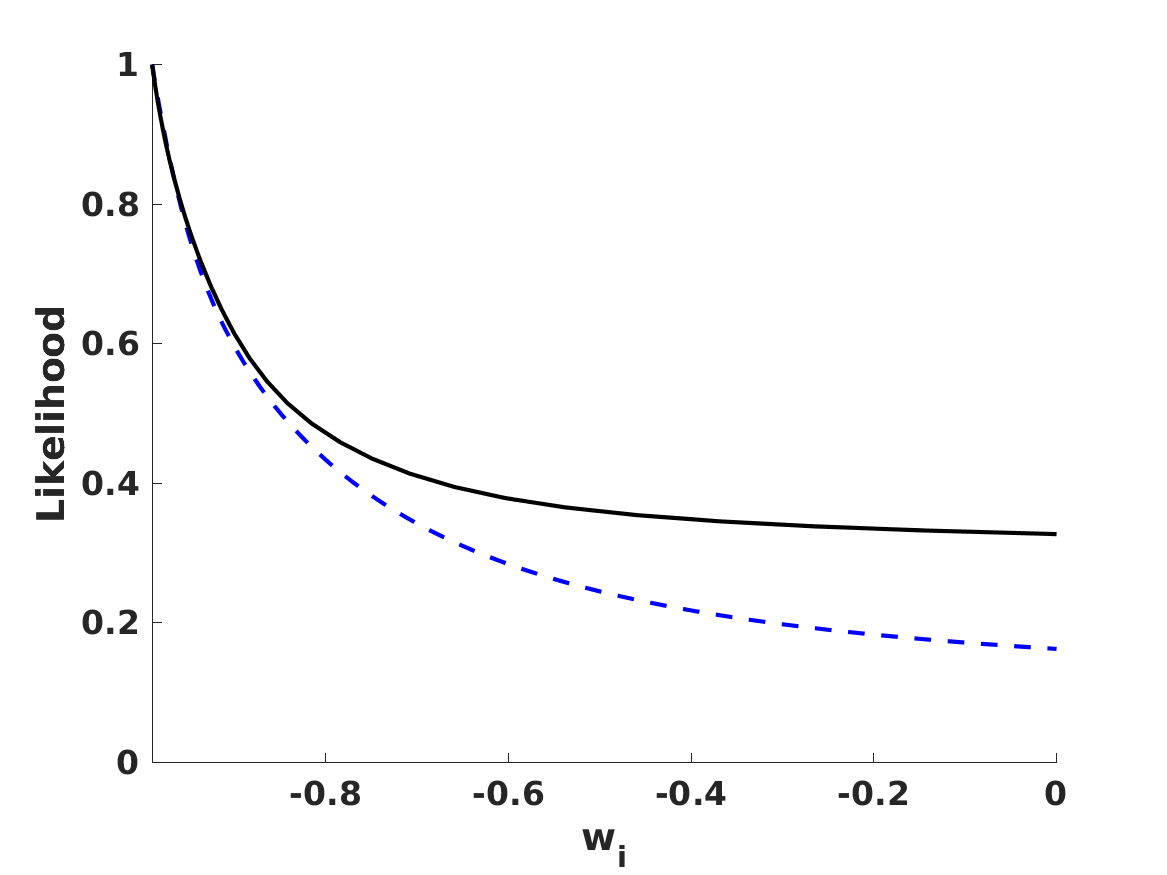}
\includegraphics[width=3.2in,keepaspectratio]{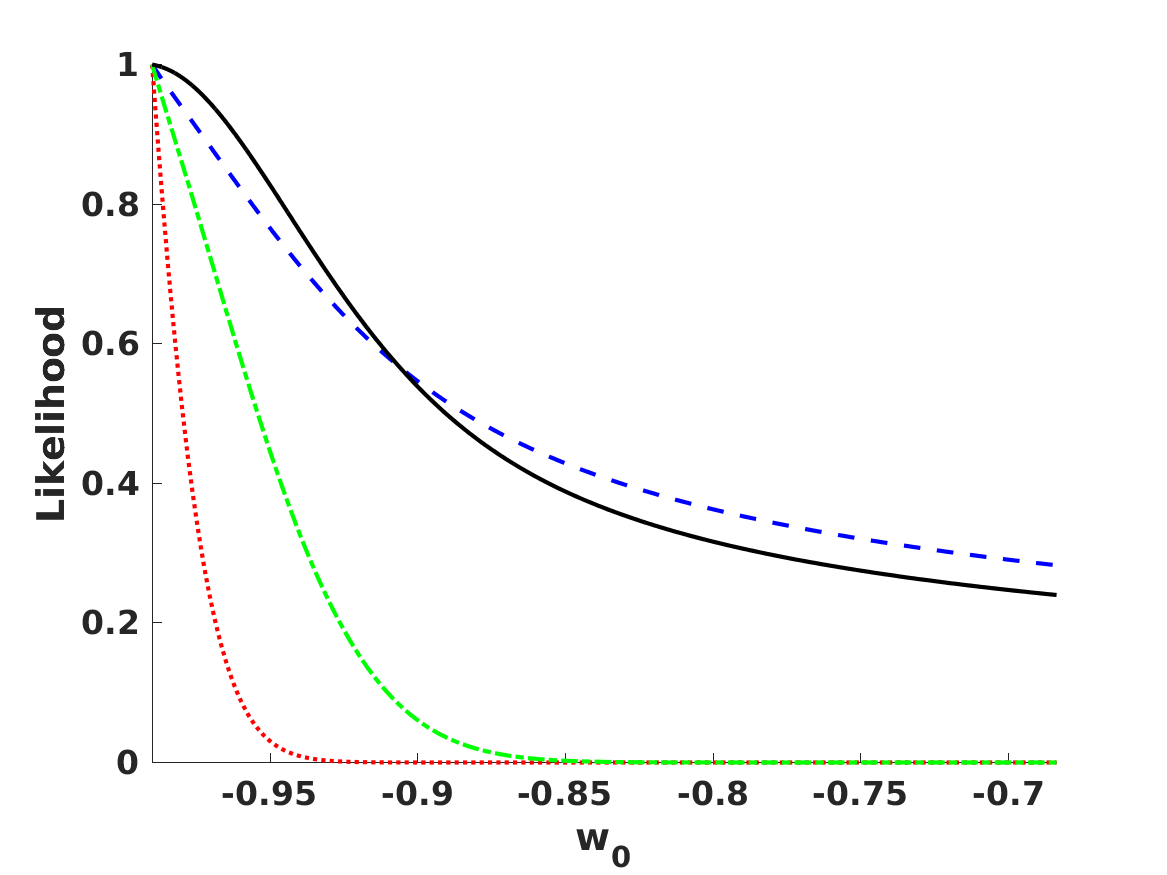}
\includegraphics[width=3.2in,keepaspectratio]{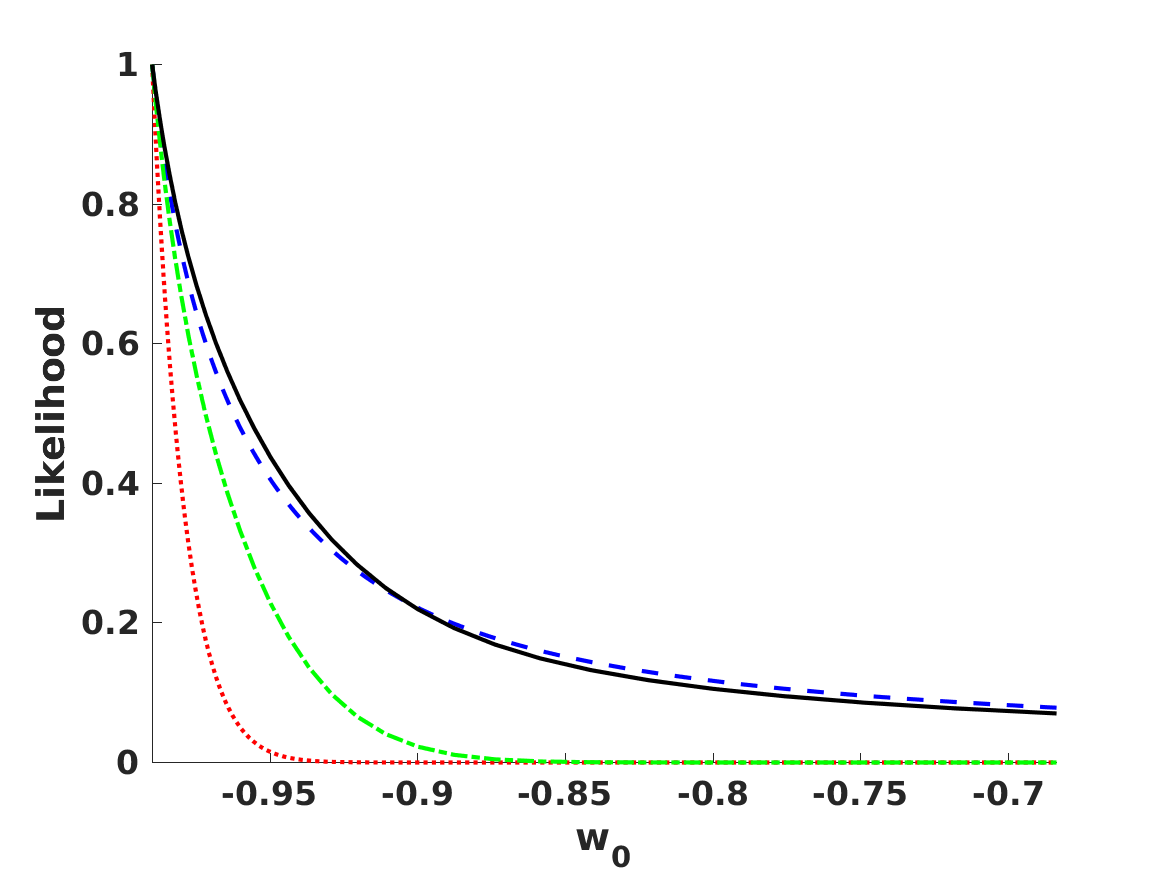}
\end{center}
\caption{\label{fig4}Comparison of the posterior likelihoods for $w_i$ (top panels) and $w_0$ (bottom panels). Left side panels show the constraints using the cosmological data only, while right side panels show the constraints using our full datasets. The red curves correspond to $w_i=0$ and $\Delta=1/3$; green curves to $w_i=0$ and $\Delta$ allowed to vary; blue curves to $\Delta=1/3$ and $w_i$ allowed to vary; black curves to the case with both parameters allowed to vary (as described in the text).}
\end{figure*}


\section{Conclusions}

We have studied a phenomenological but physically motivated class of dark energy models whose equation of state can (depending on parameter choices) undergo a rapid transition at low redshifts, and used low-redshift data to constrain it. Such a toy model is useful to quantify by how much the dark energy equation of state of dark energy may differ from a cosmological constant at redshifts of a few (given that at redshifts close to zero the answer is that the difference must be very small).

Our analysis complements and extends that of \cite{MPC} and of other recent works \cite{Lazkoz,Jaber,Durrive}. While each of these works constrains specific model classes, they confirm that any such transitions are tightly constrained. The present-day value of the dark energy equation of state, $w_0$ is always constrained to be close to that of a cosmological constant, and the same is true of $w_i$ if one allows it to vary. A summary of these constraints on $w_i$ and $w_0$ is in Figure \ref{fig4}. Thus low-redshift data pushes any such transition to high redhsifts, well into the matter era. In other words, there is no observational evidence for a transition associated with the onset of acceleration (or, for that matter, of dark energy domination, which occurs at a slightly different redshift).

The main novelty of our work consisted of constraining these dark energy transition models by a combination of cosmological data (specifically Type Ia supernova and Hubble parameter measurements), astrophysical and local measurements, exploiting the fact that realistic scalar field driven dynamical dark energy models should lead to a varying fine-structure constant. We emphasize that these constraints (especially the local ones) are competitive with the cosmological ones. A simple way to see this is that using cosmological data alone the constraints on $w_0$ and $w_i$ are comparable (cf. Table \ref{table1}) while using our full dataset the constraints on the former are clearly stronger than those on the latter (cf. Table \ref{table2}). The imminent arrival of more stringent astrophysical tests of the stability of $\alpha$ from the ESPRESSO spectrograph \cite{ESPRESSO} will enable significantly stronger constraints in the near future.

Finally, our work highlights the point that if one wants to identify possible deviations from a cosmological constant, it is important to develop observational tools that are capable of probing this behaviour in a wide redshift range, including deep in the matter era. Astrophysical tests of the stability of fundamental couplings, which we used in the present work, are one such example: measurements of $\alpha$ have already been done up to redshift $z\sim4$, and there are good prospects to extend this range to at least $z\sim6$ in the near future, which will further improve current constraints. Another possibility, that should be realized in the coming years are direct measurements of the expansion of the universe \cite{Liske,Ramos}. Quantitative studies of synergies between these observables remain to be done.

\section*{Acknowledgements}

We are grateful to Ana Catarina Leite for helpful discussions on the subject of this work. This work was financed by FEDER---Fundo Europeu de Desenvolvimento Regional funds through the COMPETE 2020---Operacional Programme for Competitiveness and Internationalisation (POCI), and by Portuguese funds through FCT---Funda\c c\~ao para a Ci\^encia e a Tecnologia in the framework of the project POCI-01-0145-FEDER-028987. M.P.C. acknowledges financial support from Programa Joves i Ci\`encia, funded by Fundaci\'o Catalunya-La Pedrera (Spain).

\bibliographystyle{model1-num-names}
\bibliography{alpha}
\end{document}